\documentclass[sigconf]{acmart}
\AtBeginDocument{%
  \providecommand\BibTeX{{%
    \normalfont B\kern-0.5em{\scshape i\kern-0.25em b}\kern-0.8em\TeX}}}

\setcopyright{acmcopyright}
\copyrightyear{2018}
\acmYear{2018}
\acmDOI{XXXXXXX.XXXXXXX}

\acmConference[ICSE 2024]{46th International Conference on Software Engineering}{April 2024}{Lisbon, Portugal}
\usepackage{multirow}

\begin{document}

\title{ReviewRanker: A Semi-Supervised Learning Based Approach for Code Review Quality Estimation}

\author{Saifullah Mahbub}
\email{saifornab@gmail.com}

\author{Md. Easin Arafat}

\author{Chowdhury Rafeed Rahman}

\author{Zannatul Ferdows}

\author{Masum Hasan}

\renewcommand{\shortauthors}{Mahbub, et al.}
\begin{abstract}
  Code review is considered a key process in the software industry for minimizing bugs and improving code quality. Inspection of review process effectiveness and continuous improvement can boost development productivity. Such inspection is a time-consuming and human-bias-prone task. We propose a semi-supervised learning based system ReviewRanker which is aimed at assigning each code review a confidence score which is expected to resonate with the quality of the review. Our proposed method is trained based on simple and and well defined labels provided by developers. The labeling task requires little to no effort from the developers and has an indirect relation to the end goal (assignment of review confidence score). ReviewRanker is expected to improve industry-wide code review quality inspection through reducing human bias and effort required for such task. The system has the potential of minimizing the back-and-forth cycle existing in the development and review process. Usable code and dataset for this research can be found at: \textit{https://github.com/saifarnab/code\_review} 
\end{abstract}




\maketitle

\section{Introduction}
The editorial world has been using peer review since 1731 \cite{spier2002history}.
Modern software development industries have given it a more common name: Code Review. Since then Modern Code Review (MCR) \cite{beller2014modern} has become an essential part of software development. MCR is a software quality control process in which one or a group of people evaluates the system by examining and analyzing different parts of source code which can be done either during or after the completion of the implementation phase. The purpose of code review is to find bugs, correct mistakes, and boost the consistency of code by improving performance and by reducing security vulnerabilities.

\begin{figure}[h]
  \centering
  \includegraphics[height=2.3cm,width=0.75\linewidth]{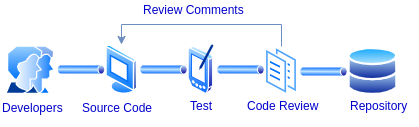}
  \caption{Code review process}
  \label{fig:code review}
\end{figure}

Figure \ref{fig:code review} outlines a typical code review process. A developer or a set of developers prepares the codes and submits them for review. A reviewer or a subgroup of reviewers then performs review checking and makes sure that the author’s codes cause no system failures in other parts of the codebase. They also ensure consistent coding style and design pattern. Following all these checking and evaluations, the reviewer or the subgroup of reviewers who have a higher role either approve or reject these reviews. Developers then make changes in codes, revise their works based on the feedback, or provide appropriate explanations against the approved review until both parties are satisfied.

 Sometimes a reviewer figures out the problematic part of the reviewed code but fails to submit an appropriate explanation of the problem. In such cases, the changes made by the developers will probably not satisfy the reviewer and we are going to get another couple of develop-review cycles. Such cycles can lead to substantial decrease in productivity in the software industry. 
 
  It is possible to minimize such situations if we can somehow assign each review a quality score. Such scoring will help us in (a) gaining a deeper understanding of quality reviews, (b) identifying quality reviewers in the company and (c) estimating provided review quality before sending off to the developers. Essentially, if after going through a particular review, a developer feels confident about the changes that he has to make in the codebase, then that review is probably of good quality. In this paper, we focus on modeling the developer confidence in a review.

  \begin{figure*}[!htb]
  \centering
\includegraphics[width=0.75\textwidth]{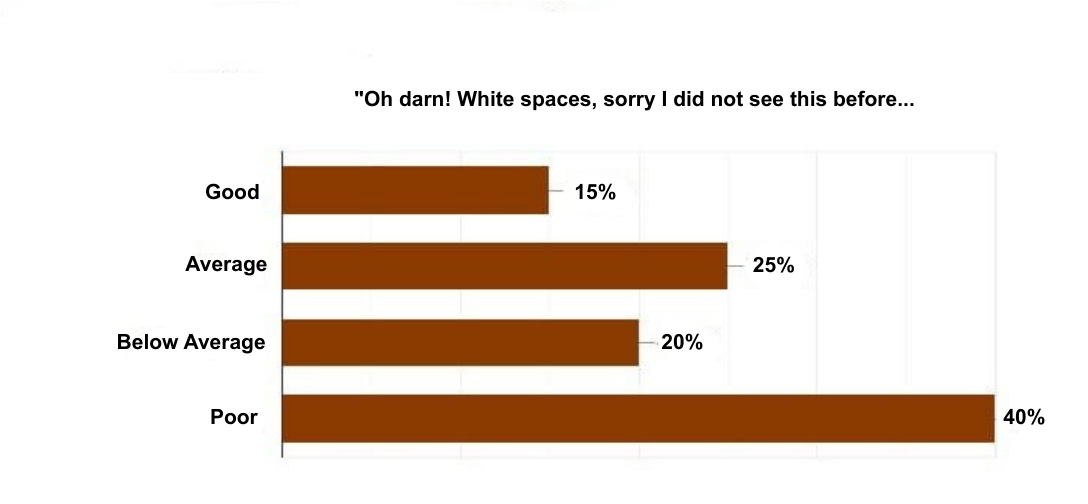}
  \caption{Human bias in code review quality labeling} 
  \label{fig:response}
\end{figure*}

One way is to simply form this task as a supervised learning task where the input will be a review and the output will be the confidence score for that review. The output labeling will be performed by the developer to whom the review had been sent for making changes in the codebase. Figure \ref{fig:response} shows the problem behind such labeling. We can see a review in the figure which has been marked as good, average, below average and poor by a significant set of developers from three different software companies. We performed this experiment on 25 reviews in total and got more or less similar results. Let us understand what this means. There are developers who are broad minded and will give good score even when the review is not that good. The opposite spectrum is also equally visible in the industry. The score assigned by a developer also depends on what type of mood he is in at that particular moment. In short, this labeling process is highly dependent on human perception which can vary widely from person to person.

We propose an alternative labeling scheme in this paper which indirectly trains a set of three models and enables them in predicting the confidence scores for a particular set of reviews. We call this semi-supervised learning approach \textit{ReviewRanker}. The labeling is related to three simple multiple choice questions (for the three models) regarding - (a) the understanding of the type of change to perform in the code, (b) the understanding of what to insert and (c) what to delete from the code based on the review of interest. We performed a similar experiment (as of Figure \ref{fig:response}) with these three multiple choice questions and found out that the choices made by the developers from different companies are similar unless the review is largely vague. Thus we have come to a conclusion that the answer to these questions are not biased by the human perception side of the developers. 

  \begin{figure*}[!htb]
  \centering
\includegraphics[width=0.6\textwidth]{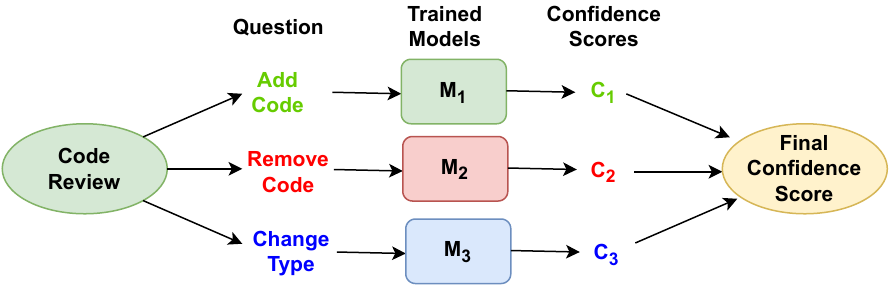}
  \caption{Code review confidence score estimation overview} 
  \label{fig:score_overview}
\end{figure*}
 
During inference (after training is done with a set of labeled reviews), we provide a code review as input to the three models for predicting the answer to the three questions (see Figure \ref{fig:score_overview}). We get three confidence scores from these three models corresponding to the ground truth answers of these questions (labeled by a developer in advance). We obtain the final confidence score from these three scores. Thus we model the confidence of the developer in understanding the review given to him or her.

Mainly three types of related studies have been performed regarding code review analysis: (1) theoretical studies on different aspects of code reviewing \cite{bosu2015characteristics,bosu2017process,kononenko2016code,spadini2018testing}, (2) assisting reviewers by problematic code snippet identification \cite{axelsson2009detecting} and (3) reviewer recommendation \cite{balachandran2013reducing,zanjani2015automatically}. Although \textbf{RevHelper} \cite{rahman2017predicting} was developed to measure code review usefulness, it is actually a binary classification tool (useful vs not useful) and does not provide any quality score to the review of interest. Also this method has the human bias aspect that we have mentioned in detail in Figure \ref{fig:response}.

\section{Problem Definition} \label{questions}
The input of ReviewRanker is a large set of code reviews $R$. The output is a confidence score $C_i$ for each review $R_i \in R$, where $C_i \in [0, 1]$. Higher confidence score denotes higher review quality. 

$C_i$ is the combination of three different confidence scores coming from three different questions related to review $R_i$. The answer of each question $Q_{ij}$ is predicted by a model $M_{j}$ that forms the question answering as a binary classification task. We get a confidence score $C_{ij}$ (associated with the ground truth label answer) from each model $M_{j}$ for each question $Q_{ij}$ for the review of interest $R_i$. The final confidence score $C_i$ of review $R_i$ is the geometric mean of all $C_{ij}$'s, where $j \in \{1,2,3\}$. 

The three questions are as follows:
\begin{enumerate}
    \item What type of operation (change in code) did the code review suggest (multi-class classification)?
    \item Did you understand what to insert in the code from the review (binary classification)?
    \item Did you understand what to delete from the code reading the review (binary classification)?
\end{enumerate}

Unlike questions related to directly assigning a quality score to a review, these three questions are straightforward and have little to no human bias.

\section{Related Works}
Researches have been undertaken to automate the process of reviewing code by using static checks such as standard violation, and common structure defects; while other researchers have focused on automating the process of reviewer recommendation and problematic code detection.

\subsection{Studies on Code Review} 
Semi-structured individual interviews were conducted with seven developers from Microsoft in \cite{bosu2015characteristics}. They concluded that prior knowledge of files leads to useful comments and tends to increase efficiency. The contemporary code review process at Microsoft was looked into in \cite{bosu2017process}. Research shows that the average spending time in a week for Microsoft developers is four hours in code review, while open source developers take five hours. Microsoft developers give more attention to reviewing relationships with developers compared to open-source developers. 
An observational survey on Mozilla’s 88 core developers was conducted in \cite{kononenko2016code}. The authors found out that approximately 57-69\% developers reviewed fewer than 5 patch files, 10\% developers reviewed 11 to 20 such files and 4\% developers reviewed more than 21 patch files each week. A study described why code review is responsible for evaluating the reliability of test codes and what professional developers do to review test codes by analyzing 300,000 code reviews from open-source projects \cite{spadini2018testing}.

\subsection{Code Review Automation Empirical Studies} 
 A prototype tool named \textbf{Code Distance Visualiser} was proposed in \cite{axelsson2009detecting} to detect problematic codes like string overflow, memory leaks, null pointer references, and incorrect API usages.  \textbf{ReviewBot} model was proposed in \cite{balachandran2013reducing} where they automated the checking for source code by using a static analyzer and recommended reviewers based on the belief that every line of code had a past history. \textbf{cHRev} model used three measurement metrics to measure the expertise of the reviewers based on their review comments: 1) higher number of review count, 2) reviewer’s effort in the workday and 3) higher weight assignment to the latest reviews \cite{zanjani2015automatically}. \textbf{RevFinder}, a recommendation model for reviewers based on file location was developed in \cite{thongtanunam2015should}. According to their heuristics, identical path files should be reviewed by identical reviewers. To analyze similar file paths, they used four string comparison techniques: 1) longest common prefix, 2) longest common suffix, 3) longest common subsequence and 4) longest common substring. \textbf{RevRec} developed in \cite{ouni2016search} consists of two models: the reviewer expertise model (RevRecRE) and the reviewer collaboration model (RevRecRC). They evaluated three open-source projects - Android, OpenStack, and Qt. A comparative study on code review usefulness was conducted based on textual features and reviewer expertise in \cite{rahman2017predicting}. The authors proposed a machine learning model named \textbf{RevHelper} to predict the usefulness of a review comment. Their comparative study was based on two heuristics - 1) differences between useful and non-useful reviews and 2) how the reviewers' experience helps them to provide appropriate reviews. 

\section{Dataset Description}
The steps regarding the dataset creation process for this research has been briefly shown in the leftmost box of Figure \ref{fig:method_overview}. We shall describe each of these steps in detail in this section.

\subsection{Data Source}
We have collected our data from multiple open-source projects hosted in Gerrit \footnote{https://www.gerritcodereview.com/}. Gerrit is a popular tool for code review in both open-source and commercial code repositories. Gerrit provides an easily accessible REST API \footnote{https://gerrit-review.googlesource.com/Documentation/rest-api.html} for collecting code reviews and their related codes. We have created a \textit{Gerrit Miner} using \textbf{Java} that mines code reviews from open source code repositories such as \textbf{Android} \& \textbf{Iotivity} and stores them in a \textbf{MySQL} database. We later query the database and label the reviews with different criteria described in detail in the upcoming subsections.

\subsection{Data Labeling} 

  \begin{figure*}[!htb]
  \centering
\includegraphics[width=0.5\textwidth]{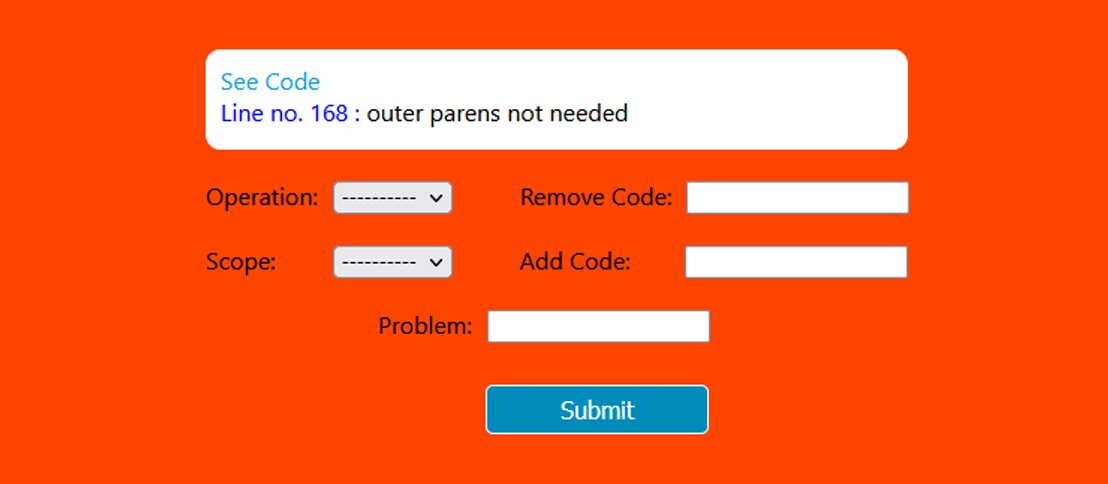}
  \caption{Data labeling app front end} 
  \label{fig:app}
\end{figure*}

  \begin{figure*}[!htb]
  \centering
\includegraphics[width=0.8\textwidth]{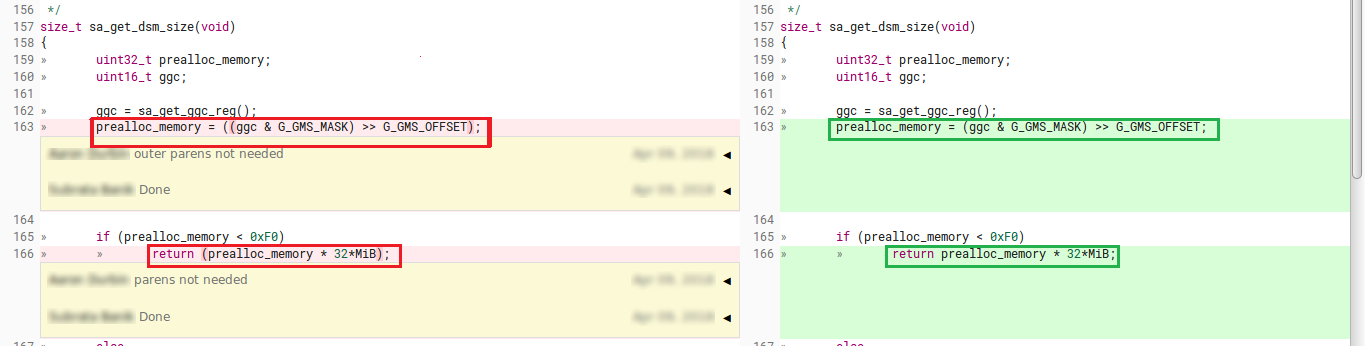}
  \caption{Code context view during code review labeling} 
  \label{fig:code_snippet}
\end{figure*}

We have created a labeling application with the \textit{Django} framework in \textbf{Python} \cite{burch2010django}.
The labeling app was designed to be user-friendly and intuitive. On entry, the web app asks for the login credentials of the user. Once it is provided, it directly goes to the labeling page and displays a code review comment to the user. The user is asked what type of operation (change type in code) the code review suggests (see Figure \ref{fig:app}). Four options are provided in the form of a drop-down menu: \textit{Insert, Delete, Replace, and Not Enough Information}. The web app provides the private URLs to the source code, and by clicking the link the user can view the source code, where the code review was submitted, and the later modification (accepted by reviewer) in the source code side by side (see Figure \ref{fig:code_snippet}). 

When the user selects one of the four operations from the drop down menu, he/she is also asked to provide the code snippet that is impacted by the operation. If the operation is an \textit{Insert} operation, the user is supposed to provide the code snippet that was to be inserted in a text field named \textit{Add Code} (only if it is understood from the review what was to be inserted). If the operation is a \textit{Remove} operation, the user puts the code that was to be removed from the original code in the text box named \textit{Remove Code} (only if it is understood from the review what was to be removed). If the operation is a \textit{Replace} operation, the user puts the part of the code that changed in \textit{Remove Code} text box, and the part that it changed into in the \textit{Add Code} text box (only if both these parts can be understood from the code review alone). We also took a human-centric design approach to design the labeling app. Each time a sample data was submitted, the web page changed the background color so that the labeling process would not become monotonous and also would give a sense of progress to the user.

\subsection{Label Validation}
The reviews were labeled by a team of five independent volunteers who possess substantial experience in programming. All the labelers are from Computer Science background and have more than two years of working experience with programming languages such as C and Java, specifically in the areas of Android and Iotivity. To ensure consistency in the labeling process, 10\% of the reviews were given to all the participants for labeling. The remaining 90\% of samples were unique for each labeler. The admin frequently examined 10\% of the data labels to check for any discrepancies among the labelers. If there was a considerable variation in the labeling, appropriate measures were taken to make the data labels more consistent. Later on, the entire dataset was manually labeled and reviewed by senior software developers to ensure proper validation of the assigned labels. The final confirmation for the labeling was obtained from the admin and considered conclusive for this dataset.

\section{Materials and Methods}
\begin{figure*}[!htb]
  \centering
\includegraphics[width=1.0\textwidth]{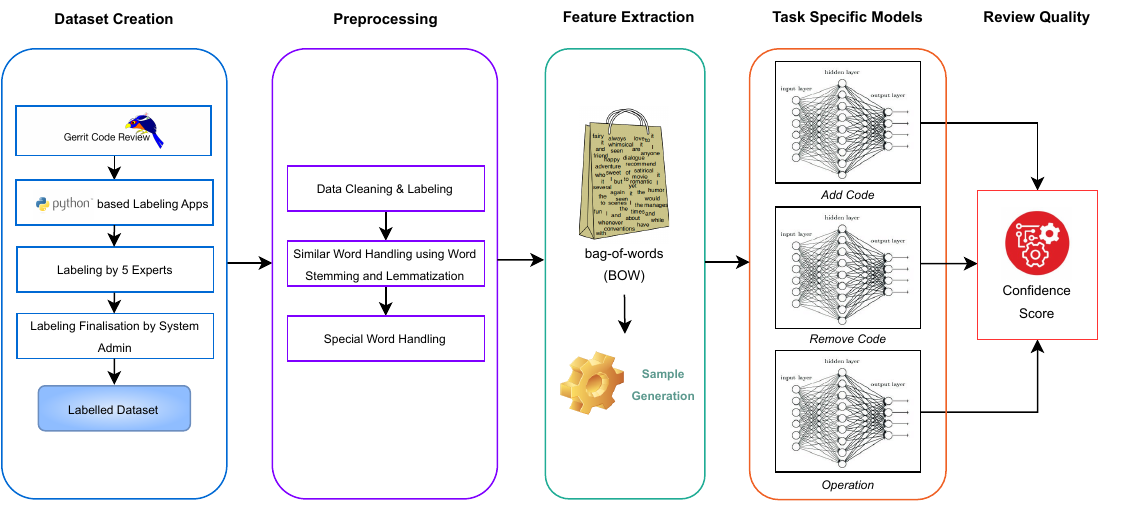}
  \caption{ReviewRanker development overview} 
  \label{fig:method_overview}
\end{figure*}

Figure \ref{fig:method_overview} provides an overview of the steps in developing ReviewRanker. We have already described the dataset creation step in the previous section. In this section, we are going to elaborate the next four steps which are more related to ReviewRanker training and inference phase. 

\subsection{Data Preprocessing}
\subsubsection{Data Labeling:} \label{data_labels}
Our initial dataset consisted of 2052 review comments. After the elimination of redundant samples, we are now left with 1483 sample reviews in our final dataset. Let us talk about the ground truth label assignment process for the three multiple choice questions asked for each review (the three questions can be found in Section \ref{questions}). In real life scenario, the ground truth labels associated to a particular review are expected to be assigned by the developer/ developers to whom the review is directed to during the development process. Observing the questions, it is evident that it will take little to no effort from the developers to perform this labeling process.

We start with the operation (code change) related question. We define four types of operations: (1) replace (class label 0), (2) delete (label 1), (3) insert (label 2) and (4) not enough information (no label assigned). If a review operation is assigned as "not enough information", then we simply assign that review a confidence score of 0 and exclude that review from ReviewRanker training and inference.

The next two questions are about understanding of what to insert and what to remove from the current code base (both are binary classification tasks). If it is clear from the review what to insert, then the insertion related question receives ground truth label of 1, else the label is 0. The exact same aspect goes for the deletion related question. 

If the operation is labeled as "replace" (first question), then it is expected that the label of both the insertion and deletion related questions will be 1 (it will not always happen in non-ideal cases). Similarly, if the operation is labeled as "delete", then the label of deletion related question is expected to be 1, while the insertion related question will have a label of 0 in an ideal world; and the opposite aspect will happen if the operation is labeled as "insert".    

Let us now look at an example review - ``outer parens not needed''. The labels for this review are as follows: \\
\textbf{Operation Type:} delete (label 1) \\
\textbf{Understanding of something to be added:} nothing to add (label 0) \\
\textbf{Understanding of something to be deleted:} parentheses need to be deleted (label 1)

\subsubsection{Similar Word Handling}
Our corpus contains more than 3000 unique words, which is a large number considering the small corpus size (less than 1500 reviews). So, by replacing all semantically identical words with a single word, we minimize the word list, which helps our model find acceptable relationships between words. While doing so, we use both the process of word stemming and lemmatization. Using word-stemming, we can modify a word’s plural instance to singular, normalize grammatical state, and so on. 
Consider the words provided below:
\begin{verbatim}
[program, programs, programmer, programming, programmers]
\end{verbatim}
\noindent The above words are generated from the word “program”. Through the word-stemming process, we replace all of these words with the word \textbf{program} in our unique word list. Using word lemmatization, we can generate a similar set of words from a single word. For example, the word \textbf{minor} generates the following words:
\begin{verbatim}
[minor, little, modest, belittled]
\end{verbatim}
\noindent These words are verbally similar to the word \textbf{minor}. Thus we replace all of these words  with the word \textbf{minor} in our unique word list as well. By doing so, our corpus now contains around 1700 unique words.

\subsubsection{Special Word Handling:} \label{final_process}
Our dataset contains code reviews that include a significant amount of special words specific to C code that have no real meaning but play a very important role in review comments. Our proposed model works based on the textual relationship between normal words and these special words. Hence we replace these words with some common words based on their operational characteristics. First, we lowercase the starting letter of all words in our corpus. After that for each of the words:

\begin{itemize}
    \item If the word has any uppercase letter, then we replace the word with \textbf{keywordvariable}, considering we usually use camel case to write variables.
    \item Otherwise, if the word contains \textbf{.h} or \textbf{\#}, then we replace the word with \textbf{keyworddoth}. The presence of such special characters denotes header files in C programming.
    \item Otherwise, if the word contains \textbf{\_}, then we replace the word with \textbf{keywordunderscore}. Having an underscore in a word is a bit confusing, it may denote a function or a variable. That is why we treat them with a special keyword.
    \item Otherwise, If the word contains parenthesis, then we replace the word with \textbf{keywordfunction}, considering all functions must initiate with a pair of parentheses.
\end{itemize}
   
 \noindent After such special keyword handling, our corpus now contains 1368 unique words which started with 3000 initially.

\subsection{Feature Extraction}
In order to feed a review to a model as input, We need a mathematical representation of that review. We have 1368 unique words in our preprocessed dataset (see Section \ref{final_process}). Each review 
contains a subset of these words. So, we represent each review with a vector $V$ of size 1368, where $V_i$ represents the total count of $word_i$ found in the review. Let us look at two examples:

\noindent \textbf{Review sample 1:} line over fifty characters you should reduce it to twenty characters.\\
\noindent \textbf{Review sample 2:} provide line level comment to line.
\\
\noindent If we create a unique word list from this corpus, it would be:
\begin{verbatim}
[line, over, fifty, characters, you, should, reduce, 
it, to, twenty, provide, level, comment]
\end{verbatim}
\noindent We can index these words from 0 to 12. The feature vector for the two sample reviews is as follows:

\begin{table}[!h]
\centering
\begin{tabular}{|c|c|c|c|c|c|c|c|c|c|c|c|c|c|}
\hline
       & 0 & 1 & 2 & 3 & 4 & 5 & 6 & 7 & 8 & 9 & 10 & 11 & 12 \\ \hline
Sample 1 & 1 & 1 & 1 & 2 & 1 & 1 & 1 & 1 & 1 & 1 & 0  & 0  & 0  \\ \hline
Sample 2 & 2 & 0 & 0 & 0 & 0 & 0 & 0 & 0 & 1 & 0 & 1  & 1  & 1 \\ \hline
\end{tabular}
\caption{Illustrative example of feature vector formation}
\label{tab:vector}
\end{table}

Instead of utilizing word embedding based approaches such as Word2Vec \cite{church2017word2vec} and FastText \cite{athiwaratkun2018probabilistic}, we have opted for a bag-of-words type of approach \cite{zhang2010understanding}. Word embedding produces semantic vectors for each word typically employed with recurrent neural networks (RNNs) \cite{medsker2001recurrent}. However, due to our small dataset and straightforward classification tasks, we have observed that a basic shallow neural network with bag-of-words feature outperforms RNNs with word embeddings through five fold cross validation.

\subsection{Model Details}

\begin{table} [!ht]
  \begin{tabular}{cl}
    \toprule
    \textbf{Model Name} & \textbf{Associated Question}\\
    \midrule
    $M_1$ & add code (insertion understanding)\\
    $M_2$ & remove code (deletion understanding)\\
    $M_3$ & operation (change type)\\
  \bottomrule
\end{tabular}
\caption{Task specific models}
\label{tab:subtask}
\end{table}

Our proposed algorithm combines three models as shown in Table \ref{tab:subtask}. Details of the classes present under each model can be found in Section \ref{data_labels}. Each model is a fully connected vanilla neural network but with a different set of parameter values. The input layer is of size 1368 (word frequency vector: total unique word no. is 1368). $M_1$ and $M_2$ are used for binary classification while $M_3$ is used for multi-class classification (three classes). \textbf{Relu} activation function \cite{nwankpa2018activation} has been used for the intermediate layers, while \textbf{Softmax} has been used for the output layer. A dropout of 20\% has been applied between each consecutive hidden layers to prevent overfitting \cite{agarap2018deep}. \textbf{Categorical Cross Entropy} \cite{zhu2018comparison} has been used as the loss function, while \textbf{Adam} (Adaptive Moment Estimation) optimizer \cite{kingma2014adam} has been used for weight update.  

\subsection{Review Confidence Score Generation}

\begin{table*}[!htb]
\begin{tabular}{|c|c|c|l|l|c|c|}
\hline
\textbf{Review}                                                                                       & \textbf{Model} & \textbf{Sub-Task} & \multicolumn{1}{c|}{\textbf{True Class}}                                      & \multicolumn{1}{c|}{\textbf{\begin{tabular}[c]{@{}c@{}}Predicted \\ Probability\end{tabular}}} & \textbf{\begin{tabular}[c]{@{}c@{}}True Class\\ Probability\end{tabular}} & \textbf{\begin{tabular}[c]{@{}c@{}}Final Confidence\\ Score\end{tabular}} \\ \hline
\multirow{3}{*}{\textbf{\begin{tabular}[c]{@{}c@{}}Line over \\ 80 characters\end{tabular}}}          & $M_1$             & Add code          & \begin{tabular}[c]{@{}l@{}}0 (add nothing\\     from review)\end{tabular}     & \begin{tabular}[c]{@{}l@{}}0: 0.973\\ 1: 0.027\end{tabular}                                    & 0.973                                                                     & \multirow{3}{*}{\textbf{0.974}}                                           \\ \cline{2-6}
& $M_2$             & Remove code       & \begin{tabular}[c]{@{}l@{}}1 (remove something\\    from review)\end{tabular} & \begin{tabular}[c]{@{}l@{}}0: 0.033\\ 1: 0.967\end{tabular}                                    & 0.967                                &                                                                           \\ \cline{2-6}
& $M_3$             & Operation         & 1 (delete)                                                          & \begin{tabular}[c]{@{}l@{}}0: 0.027\\ 1: 0.983\\ 2: 0.222\end{tabular}                         & 0.983                                                                     &                                                                           \\ \hline
\multirow{3}{*}{\textbf{\begin{tabular}[c]{@{}c@{}}Above it's \\ written as one\\ word\end{tabular}}} & $M_1$             & Add code          & \begin{tabular}[c]{@{}l@{}}1 (add something\\     from review)\end{tabular}   & \begin{tabular}[c]{@{}l@{}}0: 0.001\\ 1: 0.999\end{tabular}   & 0.999                             & \multirow{3}{*}{\textbf{0.732}}                                  \\ \cline{2-6}
& $M_2$             & Remove code       & \begin{tabular}[c]{@{}l@{}}1 (remove something\\    from review)\end{tabular} & \begin{tabular}[c]{@{}l@{}}0: 0.557\\ 1: 0.443\end{tabular}        & 0.443                                 &                              \\ \cline{2-6}
& $M_3$             & Operation         & 0 (replace)              & \begin{tabular}[c]{@{}l@{}}0: 0.888\\ 1: 0.083\\ 2: 0.029\end{tabular}                         & 0.888                  &  \\ \hline
\end{tabular}
\caption{Illustrative examples of code review confidence score generation process}
\label{confidence_generate}
\end{table*}

Table \ref{confidence_generate} illustrates the entire process of confidence score generation for two sample reviews (We assume that the three task specific models $M_1$, $M_2$ and $M_3$ are already trained).
The feature vector of each review is passed through all three models separately. Each model provides a discrete probability distribution of the task specific classes. For example, model $M_3$ always provides three probability values (sums to 1) for the three operation type specific classes. For each model, we only take the probability score associated with the ground truth class label (expected to be available for all reviews). Thus, for one review, we get total three confidence scores (predicted probability values) from the three models. The final confidence score is the geometric mean ($(C_1 \times C_2 \times C_3)^{1/3}$) of these three confidence scores. A higher confidence score denotes higher review quality, as it is expected that the developer confidence in such reviews will be high.

\subsection{Confidence Score Generation for the Entire Review Set}
The expected input to the ReviewRanker system is not a single review, but an entire set of labeled (the three questions/ tasks) reviews. The three models that are part of ReviewRanker are trained on a fraction of this labeled review set. The confidence scores for the reviews are obtained in a 10-fold cross validation style. Let us understand the entire process. Given a large set of labeled reviews $S$, we first randomly divide the set into 10 small disjoint subsets $S_1, S_2, \ldots S_{10}$ of reviews. For fold no. $i$ of the 10-fold cross validation, we use all $S_j$ ($j \neq i$) subsets of reviews for training the three models (from randomly assigned initial weights) and finally, use the trained models to predict the final confidence scores of the validation review subset $S_i$. After doing this 10 times for the 10 folds, we are going to get review confidence scores for all the reviews available in the entire review set $S$. The important thing to note here is that the confidence score of each review is obtained only when that review is part of the validation subset. This is done to avoid obtaining overfitted scores on training data (many of the confidence scores of training data are close to 1).

\section{Results and Discussion}
\subsection{Manual Inspection of Assigned Review Quality}

\begin{table*}[h]
\centering
\begin{tabular}{lll} \hline
\textbf{Review} & \textbf{Confidence Score
} & \textbf{Verdict} \\ \hline
\begin{tabular}[c]{@{}l@{}} Careful, this is a running number. No two xxx\_resource() calls may \\ have the same number there or they'll overwrite each other \end{tabular} & 0.999 & Good \\ \cline{1-3}
\begin{tabular}[c]{@{}l@{}} Could you please separate this change out into single \\ commit to keep SoC and mainboard changes separate? \end{tabular} & 0.968 & Good \\ \cline{1-3}
I'm not sure about the `-C' maybe only add the long option for now? & 0.92 & Good \\ \cline{1-3}
\begin{tabular}[c]{@{}l@{}} "That tab will most certainly elicit a response by one of the \\ update license headers" scripts. \end{tabular} & 0.86 & Good \\ \cline{1-3}
\begin{tabular}[c]{@{}l@{}}don't enable it now without adding proper ASL entries. \\ This can be push later with required ASL init \end{tabular} & 0.823 & Good \\ \cline{1-3}
\begin{tabular}[c]{@{}l@{}} While you're here could you please put this into the if C statement? \\ if (IS\_ENABLED(CONFIG\_ELOG\_BOOT\_COUNT) \&\& !s3wake) \end{tabular} & 0.746 & Average \\ \cline{1-3}
\begin{tabular}[c]{@{}l@{}} Why not just provide a define that is the C-state register? \\ \#define BLAH (ACPI\_PMIO\_BASE + 0x14) \end{tabular} & 0.717 & Average \\ \cline{1-3}
(foo*)" should be "(foo *)"" & 0.654 & Average \\ \cline{1-3}
\begin{tabular}[c]{@{}l@{}} ...and these (or are these just numbers? Well, it's unclear, so some \\ sort of documentation is needed) \end{tabular} & 0.638 & Average \\ \cline{1-3}
region\_device\_sz(\&file\_data) & 0.634 & Average \\ \cline{1-3}
nit: DDI2\_HPD\_ODL & 0.555 & Poor \\ \cline{1-3}
drop & 0.443 & Poor \\ \cline{1-3}
no \#ifs & 0.345 & Poor \\ \cline{1-3}
uintptr\_t & 0.231 & Poor \\ \cline{1-3}
212\_IRQ like line 37? & 0.189 & Poor \\ \cline{1-3}
\end{tabular}
\caption{Sample code reviews of different qualities and corresponding ReviewRanker assigned confidence scores}
\label{review_analyze}
\end{table*}

We examine both the review text and its corresponding confidence score to gain insight into the behavior of the proposed ReviewRanker system. Our goal is to understand why certain reviews receive higher scores than others. To this end, we randomly selected several reviews with high, average, and low confidence scores and analyzed their content (shown in Table \ref{review_analyze}). Through our analysis, we discovered that reviews with higher confidence scores are generally easy to understand, provide clear suggestions for changes to the code, and use specific variable and function names. Reviews with average confidence scores are sometimes easy to understand but lack substantive information, are excessively long, or contain lengthy blocks of code. Reviews with very low confidence scores are often too short to understand, lack meaningful information, and include asterisks and other special characters. Since ReviewRanker is composed of three training based neural network models, it is a data hungry system. So, larger the provided review set, better will ReviewRanker be able to model the developer confidence in a particular review.

\subsection{Model Performance}

\begin{table*}[!htb]
\begin{tabular}{|c|c|c|c|}
\hline
\textbf{Model} & \textbf{\begin{tabular}[c]{@{}c@{}}Mean Training\\ Sample No.\end{tabular}}           & \textbf{\begin{tabular}[c]{@{}c@{}}Mean Validation\\ Sample No.\end{tabular}}      & \textbf{\begin{tabular}[c]{@{}c@{}}Mean Validation\\ Accuracy\end{tabular}} \\ \hline
$M_1$             & \begin{tabular}[c]{@{}c@{}}Class 0 - 370\\ Class 1 - 805\end{tabular}                 & \begin{tabular}[c]{@{}c@{}}Class 0 - 45\\ Class 1 - 85\end{tabular}                & 98.17\%                                                                     \\ \hline
$M_2$             & \begin{tabular}[c]{@{}c@{}}Class 0 - 455\\ Class 1 - 720\end{tabular}                 & \begin{tabular}[c]{@{}c@{}}Class 0 - 55\\ Class 1 - 75\end{tabular}                & 96.71\%                                                                     \\ \hline
$M_3$             & \begin{tabular}[c]{@{}c@{}}Class 0 - 355\\ Class 1 - 375\\ Class 2 - 445\end{tabular} & \begin{tabular}[c]{@{}c@{}}Class 0 - 35\\ Class 1 - 40\\ Class 2 - 55\end{tabular} & 96.55\%                                                                     \\ \hline
\end{tabular}
\caption{Training \& validation mean sample no. and mean validation performance across the 10 folds}
\label{acc_performance}
\end{table*}

Table \ref{acc_performance} shows the dataset size and performance of the three ReviewRanker models across the 10 folds. The high mean validation accuracy shows that the models can learn to answer the three simple questions associated with review confidence score generation effectively and can generalize well to validation data. The reported performance has some implications on the usage of ReviewRanker. If for some particular set of code reviews, we see that the 10-fold cross validation performance is not upto the mark, then what it means is that the three models have not been able to understand how to answer the three questions for the provided reviews. In that case, the final confidence score provided by ReviewRanker will not be a reliable metric to measure review quality. 

\subsection{ReviewRanker Validation}
ReviewRanker has not been validated at industry-wide scale. We have made effort of validating ReviewRanker at small scale in three different software companies. But just as we have mentioned in the \textit{Introduction} section, there is high human bias when it comes to assigning some kind of quality score to a review manually as part of the labeling process. Hence, our effort remains unsuccessful. Nevertheless, this is a system that has the potential of providing us with effective review quality score at industry scale. The system works end-to-end. The input is a set of reviews (no limitation in the number of reviews provided in the set) and the output is a csv file containing confidence score for each of the provided reviews. These scores can be used to find out characteristics of high, average and poor quality reviews; which in turn can aid software industries in coming up with proper guidelines for providing code reviews. This can save considerable time and cost by minimizing the occurrence of develop-review-develop cycles. Designing an effective industry-wide validation study can be an immediate next research step for ReviewRanker.

\subsection{Limitations}
ReviewRanker asks three questions regarding change type, code addition and code deletion while providing confidence score for a particular review. It does not use the context of code based on which the review has been provided. But we firmly believe that usage of code review context by the models for answering the three questions can greatly benefit the confidence score generation process. In such a case, sequence modeling approaches such as Long Short Term Memory (LSTM) \cite{hochreiter1997long} or Transformer \cite{vaswani2017attention} can be used as the three models of ReviewRanker. But one also has to take note of the fact that these sequence models are extremely data hungry. So, if a particular review set has less than 10K reviews (which is our case as well), then it is better to use the simple feature extraction method and model architecture that we have proposed. The three questions that we ask the developers to label for each sample are not based on any large scale study. We believe that a more optimal set of questions can be used for review quality estimation provided that a well designed large scale study is undertaken for this purpose. The reviews that we are dealing with in the experimental dataset for ReviewRanker are line-level code reviews. We have not tested the method on block-level code reviews, although we expect similar result for such case as well. Finally, because of the human bias factor, proper validation of the proposed ReviewRanker method could not be performed.

\section{Conclusion}
In this paper, we propose ReviewRanker with the goal of enabling effective inspection of code review quality. We discover the human bias factor of a supervised learning based approach and thus resort to a human-bias free multiple choice question scheme in order to indirectly get the confidence score for each review in a semi-supervised fashion. We ensure that the labeling process requires little to no effort from the developers. ReviewRanker can handle a large number of reviews (theoretically no limitation in number of reviews provided) and can provide the confidence score for each review in an end to end manner with zero external effort required. The proposed system can be implemented easily at industry level to consistently identify the best reviewers and promote the best review practices with minimal time and effort. The adoption of this system is expected to enhance code quality and to reduce the back-and-forth cycle of the review process. Some immediate future research directions are - (a) well designed industry scale evaluation of ReviewRanker effectiveness in review quality estimation, (b) incorporation of code context in ReviewRanker models and (c) replacing the current set of questions with more suitable set of questions through large scale study. We plan to make ReviewRanker publicly available in the form of a Python package upon acceptance.

\bibliographystyle{ACM-Reference-Format}
\bibliography{main}

\end{document}